\documentclass[12pt,aps,preprint,pra,showpacs]{revtex4-1}
\usepackage{graphicx}

\begin{document}

\title{Prediction of a quantum anomalous Hall state in Co decorated silicene}

\author{T. P. Kaloni, N. Singh, and U. Schwingenschl\"ogl}

\email{udo.schwingenschlogl@kaust.edu.sa,+966(0)544700080}

\affiliation{Physical Science \& Engineering Division, KAUST, Thuwal 23955-6900,
Kingdom of Saudi Arabia}

\begin{abstract}
Based on first-principles calculations, we demonstrate that Co decorated silicene can host a
quantum anomalous Hall state. The exchange field induced by the Co atoms combined with 
the strong spin orbit coupling of the silicene opens a nontrivial band gap at the K-point.
As compared to other transition metals, Co decorated silicene is unique in this respect,
since usually hybridization and spin-polarization induced in the silicene suppress a quantum
anomalous Hall state. 
\end{abstract}
\maketitle

\section{Introduction}
Silicene is a single layer of Si atoms arranged in a two-dimensional honeycomb
lattice \cite{verri} and therefore closely related to graphene. It nowadays attracts
considerable attention due to its exotic electronic structure and promising
applications in Si nanoelectronics. The chemical similarity between silicene and graphene
arises from the fact that Si and C belong to the same group in the periodic table. 
However, Si is subject to $sp^3$ hybridization, whereas pure $sp^2$
hybridization is energetically favorable in graphene. Silicene has a buckled structure 
with two sublattices displaced vertically with respect to each other. 
The synthesis of silicene nanoribbons on the anisotropic 
Ag(110) surface has been studied early \cite{padova,padova1}, while
recently growth on Ag(111) has been reported \cite{vogt,feng}.
A combined experimental and theoretical study by Fleurence and coworkers \cite{antoine} on the 
formation of silicene on ZrB$_2$ thin film has indicated that the buckling is influenced
by the interaction with the substrate, leading to a direct electronic band gap at the
$\Gamma$-point. Simulations based on density functional theory show that the electronic
structure of silicene is governed by a Dirac cone, similar to graphene \cite{olle}. 
Around the Fermi energy, the charge carriers behave as massless Dirac fermions in the
$\pi$/$\pi^*$ bands, which approach each other at the K-point. The electronic
structure of silicene in a perpendicular electric field has been addressed in Ref.\ \cite{Ni}
and the results of Drummond and coworkers \cite{falko} suggest that 
variation of the electric field with respect to the strength of the spin orbit coupling 
induces a transition between a topological and a band insulator. 

Unlike the quantum Hall effect, the quantum anomalous Hall (QAH) effect results from
breaking of the time-reversal symmetry by an exchange field combined with strong spin orbit
coupling that induces a band inversion. In topological insulators the quantum spin Hall effect
has been observed experimentally \cite{science1} and the QAH effect has been predicted \cite{science2}. 
A QAH state also has been predicted by Yu and coworkers \cite{science} for magnetically
doped thin films of topological insulators and by Ezawa for silicene nanoribbons \cite{izawa}.
However, the effect has not been demonstrated for silicene so far, although transition
metal adsorption has been studied theoretically in Ref.\ \cite{PRB}, however, without
inclusion of the spin orbit coupling and onsite Coulomb interaction. Therefore, the QAH state
could not be detected in this work. In our present study, we analyze the structural, magnetic,
and electronic properties of silicene decorated by the transition metals Ti, V, Cr, Mn, Fe,
Co, Ni, or Cu in comparison to each other. Co decoration is found to result in a unique
nature, because all other elements under investigation turn out to inhibit creation of a QAH
state for different reasons.

\section{Computational details}
We perform geometry optimizations within the generalized gradient approximation including
the van der Waals interaction \cite{grime,kaloni-jmc} as implemented in the Quantum ESPRESSO
package \cite{paolo}. Moreover, we use a plane wave cutoff energy of 544 eV and a
Monkhorst-Pack $16\times16\times1$ k-mesh for $4\times4\times1$ supercells of silicene with
a lattice constant of $a=15.44$ \AA\ and a vacuum layer of 20 \AA. Our supercells contain
32 Si atoms and 1 transition metal atom, such that the density of the impurities is
low enough to neglect their mutual interaction.
The atomic positions are optimized until all forces have converged to less than 0.003 eV/\AA.
We calculate the electronic band structures of transition metal decorated silicene 
by the full potential augmented plane wave method implemented in the WIEN2k code \cite{Wien2k}. 
The application of a finite onsite Coulomb interaction $U$ is necessary
for correctly describing the $d$ electrons of the transition metal atoms, as previous studies
suggest that the adsorption geometries and electronic configurations are sensitive to
correlation effects \cite{mousumi,wehling}. A value of $U=4$ eV is expected to give appropriate
results for the transition metal atoms under consideration \cite{new3,new4} and, therefore, is
employed in our calculations. We have also tested different values of the onsite interaction
from 3 to 5 eV without finding any influence on our conclusions. The convergence threshold
is set to 10$^{-4}$ eV with $R_{mt}K_{max}=7$, where $K_{max}$ is the planewave cutoff and
$R_{mt}$ is the smallest muffin-tin radius. A k-mesh with 36 points in the irreducible
Brillouin zone is used. Silicene has been demonstrated to exist on various substrates.
While we do not take into account a specific substrate in our calculations, our results
are valid not only for a suspended sample but also in the case that the interaction with
the substrate is small.

\section{Structural Considerations}

\begin{table} [t]
\begin{tabular}{|c||c||c|c|c|c|c|c|c|c|c|}
\hline
&\begin{tabular}{c} $E_b$ top \\ (eV) \end{tabular}&\begin{tabular}{c} $E_b$ bottom \\ (eV) \end{tabular}&$h$ (\AA) &$\theta$ $(^{\circ})$&\begin{tabular}{c} Total \\ ($\mu_B$) \end{tabular} &\begin{tabular}{c} Metal $d$ \\
($\mu_B$)\end{tabular}&\begin{tabular}{c}Si \\ ($\mu_B$)\end{tabular}& \begin{tabular}{c} Interstitial \\ ($\mu_B$) \end{tabular} &\begin{tabular}{c} spin up \\ occup. \end{tabular} &\begin{tabular}{c} spin down \\ occup. \end{tabular} \\
\hline
Ti     &3.58&3.60      &1.40      &112-118    &2.05          &1.59           &0.05         &0.40        &1.90         &0.31       \\
\hline
V      &3.16&3.52      &1.36      &113-117    &2.99          &2.55           &0.02         &0.41        &2.80         &0.25         \\
\hline
Cr     &2.00&2.33      &1.30      &115-118    &4.76          &4.10           &0.08         &0.59        &4.21         &0.11     \\
\hline
Mn     &2.39&2.40      &1.10      &111-115    &3.10          &3.97           &$-$0.42      &$-$0.44     &4.40         &0.43        \\
\hline 
Fe     &2.91&3.45      &0.96      &111-118    &2.10          &2.69           &$-$0.25      &$-$0.33     &4.39         &1.70     \\
\hline  
Co     &3.42&3.99      &0.79      &113-118    &0.99          &1.16           &$-$0.07      &$-$0.10     &4.48         &3.32      \\
\hline
Ni     &3.05&3.57      &0.75      &113-117    &0.00          &0.00           &0.00         &0.00        &4.08         &4.08       \\
\hline   
Cu     &2.10&2.50      &0.66      &113-117    &0.00          &0.00           &0.00         &0.00        &4.40         &4.40     \\
\hline   
\end{tabular}
\caption{Transition metal decoration at the top site (binding energy) and hollow site (binding
energy, structural parameters, total magnetic moment, transition metal $d$ moment, 
Si moment, interstitial moment, and transition metal occupations for the spin up and down channels).}
\end{table}

\begin{figure}[t]
\includegraphics[width=0.48\textwidth,clip]{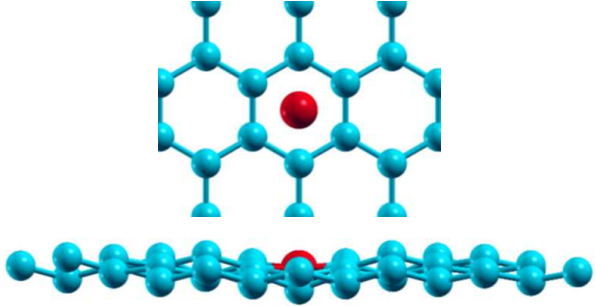}
\caption{Top and side views of the optimized structure for transition metal decoration of
silicene at the hollow site.}
\end{figure}

Because of the hexagonal symmetry of silicene, the possible decoration sites for a single atom 
can be categorized as top, bridge, and hollow. While all three choices have been studied,
decoration at the bridge site will no longer be considered in the following, because it turns
out to be an unstable configuration. The structural optimization demonstrates that a
transition metal atom added at the top site shifts close to an original Si position and
displaces the Si atom. Therefore, one metal$-$Si bond length
of 2.43 \AA\ (Ti), 2.41 \AA\ (V), 2.38 \AA\ (Cr), 2.35 \AA\ (Mn), 2.33 \AA\ (Fe), 2.30 \AA\ (Co),
2.28 \AA\ (Ni), or 2.26 \AA\ (Cu) is realized. In addition, the metal atom is bound to three
other Si atoms with shorter bond lengths of 2.39 \AA\ (Ti), 2.37 \AA\ (V), 2.35 \AA\ (Cr),
2.31 \AA\ (Mn), 2.29 \AA\ (Fe), 2.28 \AA\ (Co), 2.26 \AA\ (Ni), or 2.24 \AA\ (Cu).
The Si$-$Si bond lengths around the impurity have values of 2.27 to 2.33 \AA\ and
thus are slightly modified as compared to pristine silicene (bond length 2.27 \AA\ \cite{yao}). 
The buckling in the silicene layer amounts to 0.34 \AA\ to 0.52 \AA\ and the binding energy
to $E_b=3.58$ eV (Ti), 3.16 eV (V), 2.00 eV (Cr), 2.39 eV (Mn), 2.91 eV (Fe), 3.42 eV (Co),
3.05 eV (Ni), or 2.10 eV (Cu). Defining $h$ as the height of the transition metal atom
above the silicene plane we obtain $h=1.90$ \AA\ (Ti), 1.80 \AA\ (V), 1.70 \AA\ (Cr),
1.20 \AA\ (Mn), 1.11 \AA\ (Fe), 1.01 \AA\ (Co), 0.91 \AA\ (Ni), or 0.82 \AA\ (Cu). Moreover,
the angle $\theta$ between the Si$-$Si bonds and the normal of the silicene sheet is
in pristine silicene $\theta=116^{\circ}$ due to the mentioned mixture of $sp^2$ and
$sp^3$ hybridizations, while around the impurity a wide range of angles, 113$^{\circ}$ to
118$^{\circ}$, is realized. For decoration at the hollow site the transition metal atom
does not displace a specific Si atom but stays rather in the center of the Si
hexagon, see Fig.\ 1. It is bound to the six neighboring Si atoms with bond lengths of
2.26 \AA\ to 2.35 \AA. The obtained values of $h$ are listed in Table I. The buckling,
the Si--Si bond length, and the angle $\theta$ are found to be slightly modified as compared
to decoration at the top site.

The energy difference between the configurations with the transition
metal atom at the top and hollow sites is 0.02 eV (Ti), 0.36 eV (V), 0.33 eV (Cr),
0.01 eV (Mn), 0.54 eV (Fe), 0.57 eV (Co), 0.52 eV (Ni), and 0.40 eV (Cu),
indicating that decoration at the hollow site is always energetically favorable.
These values are in reasonable agreement with previous results on transition metal decorated
silicene \cite{PRB}, which also applies to the optimized structures. Only the heights $h$ are
significantly different, which we attribute to the inclusion of the van der Waal interaction 
in our structural optimizations and the spin orbit coupling in our electronic structure
calculations. It thus is to be expected that our results are more reliable than the previously
reported values. 
  
\section{Electronic structure}
In the analysis of the electronic structure, we consider only the energetically favorable
hollow site, for the different transition metal impurities. In the case of Ti, V, and Cr
decoration we obtain total magnetic moments of 2.05 $\mu_B$, 2.99 $\mu_B$, and 4.10 $\mu_B$
per unit cell, respectively, see Table I. It should be noted that the main portion
to the magnetic moment comes from the transition metal $d$ orbitals with small contributions
form Si atoms and the interstitial region. We note that the silicene sheet gets significantly
polarized, where the magnetic moments of the Ti, V, and Cr atoms are aligned 
ferrimagnetically with respect to the induced Si spins. For half filling (Mn)
and beyond half filling (Fe and Co) the orientations of the transition metal and Si spins are
opposite (antiferromagnetically aligned; negative signs in Table I).  
In contrast to Mn and Fe decoration, in the case of Co decoration the magnetic moment
of 0.99 $\mu_B$ is strongly localized and almost completely due to the Co $d$ orbitals.
This fact implies that Co affects the silicene sheet much less than the other transition 
metal atoms with larger local magnetic moments. Finally, silicene decorated with Ni or Cu
turns out to be non-magnetic, as to be expected.  

\begin{figure}[t]
\includegraphics[width=0.8\textwidth,clip]{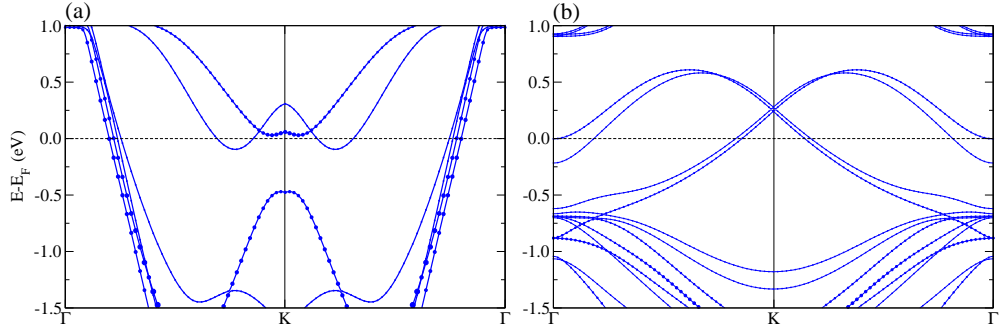}
\caption{Electronic band structure with weights of the Co 3$d$ states (size of the dots)
for decorated silicene (a) $1\times$1 and (b) $2\times2$ supercells.}
\end{figure}

We consider different coverages of transition metal atoms by employing $1\times1\times1$, 
$2\times2\times1$, $3\times3\times1$, and $4\times4\times1$ supercells of silicene to
which a single transition metal atom is added. We find in none of the supercells
besides the $4\times4\times1$ supercell signs of a QAH state, indicating that
interaction between the transition metal atoms counteracts the creation of this state
and that, thus, the impurity density has to be sufficiently low. In the case of a high
transition metal coverage, see Fig.\ 2(a,b) for the example of Co decoration, the
Co$-$Co interaction, coming along with the small separation of only 3.86 \AA\ (periodic
boundary conditions), modifies the shape of the band structure in the vicinity of the
Fermi energy strongly. For different coverages accordingly fundamentally different band
structures are obtained. On the other hand, the distance between the transition metal atoms
in the $4\times4\times1$ supercell is large enough to represent the dilute limit. A close
similarity of the $4\times4\times1$ and $5\times5\times1$ band structures indicates that
already in the former case the limit of low impurity density is reached. In the following we 
address the $4\times4\times1$ supercell for this reason.

In the band structures of Ti decorated silicene, see Fig.\ 3(a), and V decorated silicene,
see Fig.\ 3(b), a strong hybridization between the transition metal 3$d$ and Si 3$p$ states
is observed. Similar findings previously have been reported for Au and Mo
doped graphene \cite{new1,new2}. Due to the hybridization the QAH effect cannot be realized
for Ti and V decoration as the silicene states are perturbed. As a consequence of the
relative energetic shift between the impurity and silicene states when another transition
metal from the same period of the periodic table is chosen, we can expect that exchange
of the impurity can overcome the hybridization problem in the vicinity of the Fermi energy.
In the case of Cr decoration, see Fig.\ 3(c), we observe the remainder of a Dirac
cone with only small transition metal weights, which demonstrates that the states near the Fermi energy
are mainly due to the Si $p_z$ orbitals. However, in this case the local magnetic moment
of the impurity is high, see Table I, such that the silicene sheet becomes to some
extent spin polarized, which also prevents the creation of a QAH state. The band structure of
Mn decorated silicene is addressed in Fig.\ 3(d). The remainder of a Dirac cone is visible about
0.25 eV below the Fermi energy, reflecting $n$-doping. At the K-point the bands are
due to the Si $p_z$ states without hybridization with the Mn 3$d$ states. Otherwise the
situation is very similar to the Cr case, except for the antiferromagnetic polarization of
the silicene as mentioned before. In Fig.\ 3(e) we deal with Fe decorated silicene. The
Dirac cone now is located about 0.12 eV below the Fermi energy at the 
high symmetry K-point, i.e., it is $n$-doped. The Fe $3d$ states likewise do not hybridize
with the Si 4$p_z$ states in the vicinity of the Fermi energy. However, similar to Mn
decoration, the high Fe magnetic moment induces significant spin polarization in the 
silicene, such that no QAH state is realized. 

\begin{figure}[t]
\includegraphics[width=0.94\textwidth,clip]{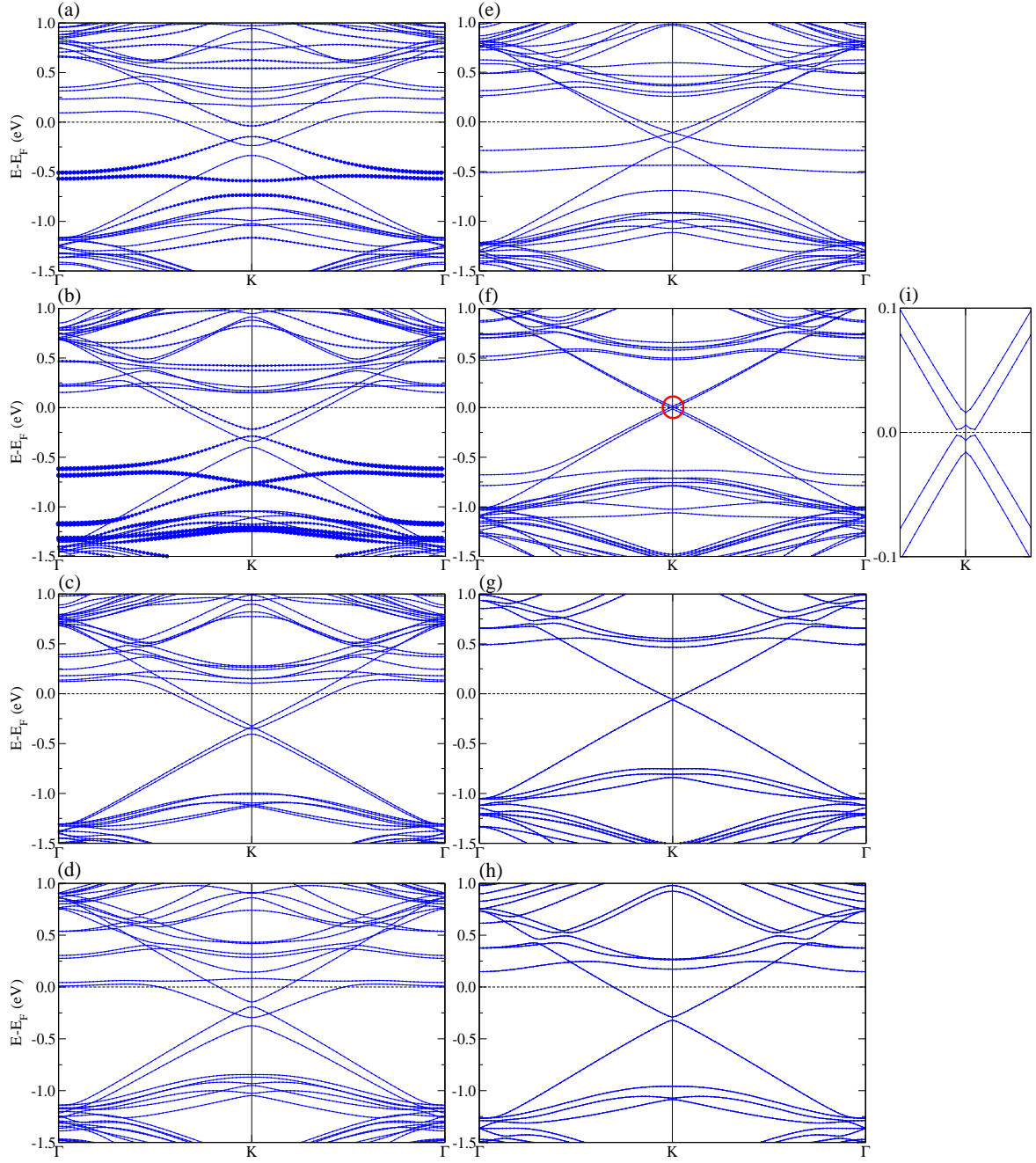}
\caption{Electronic band structure with weights of the transition metal 3$d$ states
(size of the dots) for (a) Ti, (b) V, (c) Cr, (d) Mn, (e) Fe, (f) Co, (g) Ni, and (h) Cu
decorated silicene (at the hollow site). (i) Zoom of the marked region of panel (f).}
\end{figure}

The electronic band structure obtained for Co decorated silicene is shown in Fig.\ 3(f).
We note a tiny energy gap at the K-point and the typical band alignment of a QAH system
\cite{science}, where the Dirac cone is centered at the Fermi energy. The exchange field
due to the Co local magnetic moment breaks the time-reversal symmetry and combined with
the strong spin orbit coupling this results in a QAH state in Co decorated silicene.
From the weighted band structure it is clear that the Dirac cone is essentially purely
due to Si $p_z$ orbitals. Therefore, the characteristic silicene states are maintained 
and the way is paved to the QAH state. The magnetic moment of Co is minimal but finite
and therefore just right to break the time-reversal symmetry, while maintaining the electronic
structure specific to silicene. Our results clearly demonstrate that a QAH state
by transition metal decoration is a rare phenomenon, as in all other cases besides Co
the $d$-$p_z$ hybridization and/or the induced spin polarization of the Si $p_z$ electrons
perturbs the electronic states. Finally, we mention that for Ni and Cu decorated silicene
non-magnetic states are obtained. Interestingly, in both these systems a Dirac cone is
observed below the Fermi level, which could by shifted back by means of a gate voltage.
Still, a QAH state could not be induced, since no exchange field is left.  

\section{Conclusion}
In conclusion, using density functional theory, we find that silicene decoration by the
transition metal atoms Ti, V, Cr, Mn, Fe, Co, Ni, and Cu results in occupation of the
hollow site of the Si honeycomb. While pristine silicene is subject to spin degeneracy,
transition metal decoration can induce substantial spin polarization, that is understood
from atomic considerations. We demonstrate that Co decorated silicene is a hybrid material
that hosts a QAH state. In the cases of Ti
and V decoration a strong hybridization of the transition metal 3$d$ states with the Si 3$p_z$
states suppresses this state. Because of the large local magnetic moments induced by
Mn and Fe the silicene in these cases becomes spin polarized, which also prohibits
the formation of the QAH state. Cr is probably affected by a combination of the two effects,
both being weaker but together enough to perturb the silicene electronic structure
sufficiently. Ni and Cu decorated silicene are found to be non-magnetic and 
therefore also not suitable for our purpose. We have demonstrated that realization
of a QAH state is possible in Co decorated silicene, as long as the Co atoms do not cluster.
On the other hand, both hybridization and induced spin polarization typically destroy the
characteristic electronic structure of pristine silicene and therefore exclude a QAH state
by transition metal decoration. Only in the case of Co decoration a sufficient but not too
large exchange field is achieved which can interact with the strong spin orbit coupling in silicene.
 
\begin{acknowledgments}
We thank M.\ Tahir for fruitful discussions and KAUST IT for providing computational resources.
\end{acknowledgments}

\end{document}